\documentstyle[twoside,fleqn,espcrc2,epsf]{article}

\title{
Improved Error Estimate for the Valence Approximation}

\author{W.\ Lee \thanks{present address:
Group T-8, Los Alamos National Laboratory, Los Alamos, NM 87545}
and D.\ Weingarten\\
IBM Research, P.O.~Box 218,
Yorktown Heights, NY 10598\\}

\begin{document}

\begin{abstract}

We construct a systematic mean-field-improved coupling constant and
quark loop expansion for corrections to the valence (quenched)
approximation to vacuum expectation values in the lattice formulation of
QCD.  Terms in the expansion are evaluated by a combination of weak
coupling perturbation theory and a Monte Carlo algorithm.
 
\end{abstract}

\maketitle

The valence (quenched) approximation to the infinite volume, continuum
limit of lattice QCD gives values for hadron
%masses~\cite{Butler_mass,MILC,CPPACS} and for meson decay
masses and for meson decay
%constants~\cite{Butler_decay,CPPACS} not far from experiment.  Missing
constants not far from experiment.  Missing
from these calculations, however, is an independent theoretical estimate
of the error arising from the valence approximation.  A possible method
for determining the valence approximation's error is proposed in
Ref.~\cite{Sexton}. In the present article, we describe an improved
version of of the method of Ref.~\cite{Sexton} which we believe is
likely to require less computer time.  The expansion we describe can be
adapted to any choice of quark action but will be given here only for
Wilson quarks.

In an exact treatment of QCD, virtual quark-antiquark pairs produced by
a chromoelectric field reduce the field's intensity by a factor which
depends both on the field's momentum and on its intensity.  In the
valence approximation this factor, analogous to a dielectric constant,
is approximated by its zero-field-momentum zero-field-intensity
limit~\cite{DW82}.  Our expression for the error in the valence
approximation to any vacuum expectation value consists of a sum indexed
by a power of a mean-field-improved coupling constant~\cite{Lepage} and
by a quark loop count.  Each term in the expansion requires
as input, in effect, the dielectric constant entering the valence approximation.
The dielectric constant, for convenience, we obtain analytically from
mean-field-improved perturbation theory to second order in the coupling
constant. A related calculation without mean-field improvement is
described in Ref.~\cite{Hasenfratz}.  The remaining work of evaluating
each term in the error expansion is done by a Monte Carlo algorithm.

In a simple test case, our error estimate requires significantly less
computer time than a direct comparison of full QCD and the valence
approximation. Whether this gain holds also for more interesting cases
we do not yet know.

We consider Wilson's formulation of euclidean QCD on some finite
lattice, with periodic boundary conditions for gauge links $u(x,y)$, and
antiperiodic boundary conditions for the coupling matrix $M$ for a
single quark flavor.  For $n_f$, either even or odd, degenerate flavors
of quarks and any function of the gauge fields $G$, the vacuum
expectation value obtained after integrating out quark fields is
\begin{eqnarray}
\label{expG}
< G > & = & Z^{-1} \int d \nu \, G \, det( M)^{n_f} \, 
          exp( \frac{\beta}{6} P ), \nonumber \\  
Z & = & \int d \nu \, det( M)^{n_f} \, 
          exp( \frac{\beta}{6} P ), 
\end{eqnarray}
where, as usual, $P$ is Wilson's plaquette action,
$\beta$ is $6/g^2$ for bare gauge coupling constant $g$, and
$\nu$ is the product of one copy of $SU(3)$ Haar measure for each link
variable on the lattice. We ignore for simplicity vacuum expectation values
of products of quark and antiquark fields.

The valence approximation to $< G>$ of Eq.~(\ref{expG}) is
\begin{eqnarray}
\label{expGv}
< G >_v & = & Z_v^{-1} \int d \nu \, G \,  
          exp( \frac{{\beta}_v}{6} P ), \nonumber \\  
Z_v & = & \int d \nu \, 
          exp( \frac{{\beta}_v}{6} P ), 
\end{eqnarray}
with valence approximation $\beta_v$ given by $6/{g_v}^2$ for valence
approximation bare gauge coupling $g_v$. It is convenient to define 
$\Delta \beta$ as the shift ${\beta}_v - \beta$.

We now derive a coupling constant and quark loop expansion for the
difference between a full QCD vacuum expectation value $< G>$ and its
valence approximation $< G>_v$.
For this purpose, we recast $< G>$ of Eq.~(\ref{expG}) as
\begin{eqnarray}
\label{expGx}
< G > & = & Z^{-1} \int d \nu \, G \,  
          exp( \frac{{\beta}_v }{6} P + Q) \nonumber \\
Z & = & \int d \nu \,  exp( \frac{{\beta}_v }{6} P + Q ), \\
Q & = & n_f tr {\log}(M) - \frac{\Delta \beta}{6} P, \nonumber
\end{eqnarray}
then introduce a parameter $\lambda$ multiplying $Q$, expand $< G>$ in
powers of $\lambda$, and replace $\lambda$ by 1. We obtain
\begin{eqnarray}
\label{errG}
\lefteqn{< G > = < G>_v +} \\ 
& & < (G - <G>_v)(Q - <Q>_v)>_v + \ldots \nonumber 
%& & < ( G - < G>_v) (Q - < Q>_v)^2 >_v \ldots \nonumber
\end{eqnarray}
The $n$th term in Eq.~(\ref{errG}) is the $n - 1$ quark loop correction
to the valence approximation. As an error estimate for the valence
approximation, we will concentrate on the one-loop correction $< (G -
<G>_v)(Q - <Q>_v)>_v$.

For $tr {\log}(M)$ in the one-loop term of Eq.~\ref{errG}, we construct
a coupling constant expansion. To do this, transform each gauge
configuration to the euclidean lattice version of Landau gauge by
maximizing $\sum_y tr[u(x,y)]$ at each lattice site $x$.  One algorithm
for finding this transformation is discussed in Ref.~\cite{Sexton}.  We
obtain some random sample of Gribov copies the details of which play no
explicit role in our expansion. For each fixed gauge configuration, let
$z$ be the average over all lattice links of $tr[u(x,y)]/3$. Let $M_0$
be a free hopping matrix with hopping constant $\kappa_0$ chosen to give
a quark mass $1/(2\kappa_0) - 4$ that agrees with a
mean-field-improved~\cite{Lepage} value of the quark mass $1/(2 z
\kappa) - 1/(2 z \kappa_c)$ carried by $M$, where $\kappa$ and
$\kappa_c$ are, respectively, the hopping constant of $M$ and the
valence approximation to the critical value of this hopping constant.

We now express $tr {\log}(M)$ in the form
\begin{eqnarray}
\label{Mexp1}
\lefteqn{tr {\log}(M) =} \nonumber \\
& & tr {\log}\{ z M_0[ 1 - M_0^{-1}(M_0 - z^{-1}M)]\}, 
\end{eqnarray}
and expand to obtain a sum indexed, in effect, 
by powers of a mean-field-improved quark coupling constant~\cite{Lepage}  
\begin{eqnarray}
\label{Mexp2}
\lefteqn{tr {\log}(M) = tr {\log}( z M_0) - } \\
&  & \sum_n \frac{1}{n} tr\{[M_0^{-1}(M_0 - z^{-1} M)]^n\}. \nonumber
\end{eqnarray}
The trace required in the second term of Eq.~(\ref{Mexp2}) we obtain
from an ensemble of complex-valued quark fields $\phi^r$, $1 \le r \le
R$ with each component of each field an independent random complex
number on the unit circle. We obtain
\begin{eqnarray}
\label{Mexp3}
\lefteqn{tr {\log}(M) = tr {\log}( z M_0) - } \\
&  & R^{-1} \sum_{nr} \frac{1}{n} < \phi^r|[M_0^{-1}(M_0 -
z^{-1} M)]^n \phi^r>, \nonumber
\end{eqnarray}
where $< \ldots | \ldots >$ is the inner product on the 
space of complex-valued quark fields. The term $tr {\log}( z M_0)$ in
Eq.~(\ref{Mexp3}) can be computed by diagonalizing $M_0$ in momentum
space, and multiplication by $M_0^{-1}$ can be done efficiently using
fast Fourier transformations.

Combining Eqs.~(\ref{errG}) and (\ref{Mexp3}) gives a coupling constant
and quark-loop expansion for corrections to valence approximation
vacuum expectation values. Still to be determined is the shift
$\Delta \beta$. Eqs.~(\ref{errG}) and (\ref{Mexp3}) are formally correct
for any choice of $\Delta \beta$. The rate at which these expansions
converge, however, will be affected by the choice of $\Delta \beta$.
One possible way to choose $\Delta \beta$ is to require the error or
some approximation to the error in a particular physical quantity to be
zero.  
For the present discussion, we choose $\Delta \beta$ by requiring that
(analytic) mean-field-improved perturbation theory give zero valence
approximation error, to first order in quark loops and second order in
the coupling constant, for the Landau gauge gluon propagator at minimal
nonzero momentum.
We obtain
\begin{eqnarray}
\label{deltabeta}
\Delta \beta  & = &  
\frac{9 n_f}{4 \sin^2(\pi/L) <tr U >_v} \times \nonumber \\
& & [\Pi_{2 2}(p) - \Pi_{2 2}(0)] \\
\Pi_{\mu \nu}(p) & = &
\frac{1}{L^4} \sum_q tr[ \Gamma_{\mu}(q + p/2 ) S(q + p) \times \nonumber \\ 
& & \Gamma_{\nu}(q + p/2 ) S(q) ] \nonumber,
\end{eqnarray}
where $L$ is the lattice period, 
$p$ is a momentum vector with a single nonzero component $p_1$ of
$2 \pi/L$, 
$<tr U>_v$ is the valence approximation plaquette expectation value, and
each component of $q$ in the sum over $q$ ranges from $\pi/L$ to 
$2\pi - \pi/L$ in steps of $2\pi/L$. The propagator 
$S(q)$ has inverse $S(q)^{-1}$ of
$1/(2 \kappa_0) - i \sum_{\mu} \gamma_{\mu} sin( q_{\mu}) - 
\sum_{\mu} cos( q_{\mu})$, and the vertex
$\Gamma_{\mu}(q)$ is $sin( q_{\mu}) - i\gamma_{\mu} cos(q_{\mu}) $.
The limiting value of $\Delta \beta$ for large $L$ without mean-field
improvement has been derived in Ref.~\cite{Hasenfratz}.

As a test of our method we compared valence approximation expectations $<
G>_v$, Eq.~(\ref{expGv}), their one-loop errors $< (G - <G>_v)(Q - <Q>_v)>_v$,
Eq.~(\ref{errG}), and the corresponding full QCD expectations $<G>$,
Eq.~(\ref{expG}), for a lattice $10^4$ with $\beta_v$ of 5.679, $\kappa$
of 0.16 and $n_f$ of 2. According to Eq.~(\ref{deltabeta}), $\Delta
\beta$ is then 0.243 giving a full QCD $\beta$ of 5.436. For this case
$\Delta \beta$ found by the method of Ref.~\cite{Sexton} is 0.244(6).
We used 224 independent equilibrium gauge configuration in the valence
approximation ensemble, generated by an over-relaxed pseudo heat bath
algorithm, 600 random fermion fields $R$ to evaluate the trace in
Eq.~(\ref{Mexp3}) and 176 weakly correlated equilibrium gauge configurations
for the full QCD ensemble, generated by a red-black preconditioned
hybrid Monte Carlo algorithm. The expansion in Eq.~(\ref{Mexp3}) was
carried to order $n$ of 10.
The calculation of $< (G - <G>_v)(Q - <Q>_v)>_v$  was not
turned carefully. In particular $R$ of 600 in Eq.~(\ref{Mexp3}) is much larger than its
optimal vaule. The time required for the valence approximation and error
calculation was still less than 5\% of the time required by the full QCD
calculations.

\begin{figure}
\epsfxsize=63mm
\epsfbox{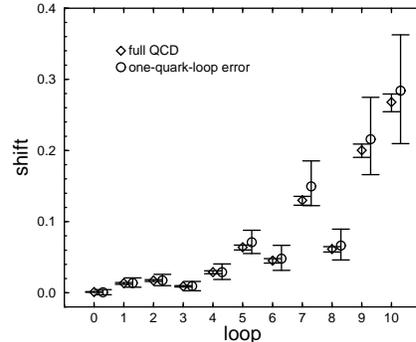}
\vskip -15mm
\caption{The predicted relative shift of 11 Wilson loops from
their valence approximation values in comparison to the true shift of
full QCD.}
\vskip -8mm
\label{fig:order7}
\end{figure}

For $G$ we used Wilson loops $W_0, \ldots W_{10}$ consisting,
respectively, of paths $1 \times 1$, $2 \times 1$, all rotations of
steps in the directions $\hat{1},\hat{2},\hat{3},-\hat{1},-\hat{2},
-\hat{3}$, all rotations of steps in the directions
$\hat{1},\hat{2},\hat{3},-\hat{2},-\hat{1}, -\hat{3}$, $3 \times 1$, $2
\times 2$, $4 \times 1$, $5 \times 1$, $3 \times 2$, $4
\times 2$, and $3 \times 3$. 
For all loops, we found that the predicted error $< (W_i - <W_i>_v)(Q - <Q>_v)>_v$
converges as a function of the highest power $n$ of coupling strength used in
Eq.~(\ref{Mexp3}) by $n$ of 7 or smaller.
For $n$ of 7, Fig.~\ref{fig:order7} shows 
the relative shift of the valence
approximation from full QCD $(<W_i> - <W_i>_v)/<W_i>$ and the
predicted value $< (W_i - <W_i>_v)(Q - <Q>_v)>_v/<W_i>$.
To within statistical uncertainties, the predicted errors agree with the
true errors. 

The true errors in Fig.~\ref{fig:order7} were found from the shortest
full QCD run sufficient to confirm equilibration of $<W_0>, \ldots
<W_{10}>$. Nonetheless the statistical uncertainties in the predicted
errors are much larger than those in the true errors.  If we were to run
the error prediction algorithm long enough to obtain statistical
uncertainties comparable to the uncertainties found by a direct comparison
of full QCD and the valence approximation, it is possible that the
computer time required by the error algorithm would become comparable to
that for full QCD. To find the uncertainty arising from use of the
valence approximation, however, the statistical uncertainty in the error
estimate does not need to be too much smaller than the error estimate's
central value. Used in this way, for the set of parameters of the test,
our algorithm takes significantly less time than the shortest possible
direct comparison of the valence approximation and full QCD.

\end{document}